\documentclass[aps,superscriptaddress,showpacs,prd,twocolumn]{revtex4-2}
\usepackage{amsfonts}
\usepackage{amssymb}
\usepackage{amsmath}
\usepackage{dsfont}
\usepackage{color}
\usepackage[dvipsnames]{xcolor}
\usepackage{float}
\usepackage{accents}
\usepackage{graphicx}
\usepackage{epstopdf}
\usepackage{ulem}
\usepackage[colorlinks=true,pdfstartview=FitV,linkcolor=blue,citecolor=blue,urlcolor=blue,breaklinks=true]{hyperref}
\usepackage{eurosym}
\usepackage{bm}
\usepackage{float}
\usepackage{wasysym}
\usepackage{blindtext}
\usepackage{longtable}
\usepackage{dcolumn}

\RequirePackage{color}

\begin{document}

\title{Restricted baby Skyrme-Maxwell theory in a magnetic medium: BPS configurations and some properties}
\author{J. Andrade}
\email{joao.luis@discente.ufma.br}
\author{Rodolfo Casana}
\email{rodolfo.casana@ufma.br}\email{rodolfo.casana@gmail.com}
\affiliation{Departamento de F\'{\i}sica, Universidade Federal do Maranh\~{a}o,
65080-805, S\~{a}o Lu\'{\i}s, Maranh\~{a}o, Brazil.}
\author{E. da Hora}
\email{carlos.hora@ufma.br}\email{edahora.ufma@gmail.com}
\affiliation{Coordena\c{c}\~{a}o do Curso de Bacharelado Interdisciplinar em Ci\^{e}ncia e Tecnologia,\\
Universidade Federal do Maranh\~{a}o, {65080-805}, S\~{a}o Lu\'{\i}s, Maranh\~{a}o, Brazil.}
\author{{A. C.} Santos}
\email{andre.cs@discente.ufma.br}
\affiliation{Departamento de F\'{\i}sica, Universidade Federal do Maranh\~{a}o,
65080-805, S\~{a}o Lu\'{\i}s, Maranh\~{a}o, Brazil.}

\begin{abstract}
We study the existence of BPS configurations in a restricted baby Skyrme-Maxwell enlarged via the inclusion of a nontrivial magnetic permeability. In order to attain such a goal, we use the Bogomol'nyi-Prasad-Sommerfield prescription, which allows us to obtain the lower bound for the energy and the BPS equations whose [electrically neutral] solutions saturate that bound. During the energy minimization procedure, we find a differential constraint which involves the self-dual potential, the superpotential itself and also the magnetic permeability. In order to solve the BPS system, we focus our attention on those solutions with rotational symmetry. For that, we fix the magnetic permeability and select two BPS potentials which exhibit a similar behavior near to the vacuum. We depict the resulting profiles and proceed to an analytical description of the properties of the BPS magnetic field. Furthermore, we consider some essential aspects of our model, such as the conditions for the overall existence of the BPS solutions, and how the permeability affects the magnetic flux. Finally, we present a family of exact BPS solutions.
\end{abstract}

\maketitle

\section{Introduction \label{sec1}}

\label{Intro}

Topologically nontrivial structures are commonly described by means of
those\ time-independent solutions which come from highly nonlinear
Euler-Lagrange equations {\cite{n5}}. In such a context, the potential term
which defines the vacuum manifold of the respective theory not only
introduces the nonlinearity itself, but it is also expected to allow the
spontaneous symmetry breaking mechanism to occur (whose effects include the
formation of a topological profile as a result of the corresponding phase
transition). {The point is that highly nonlinear equations of motion are
typically quite hard to solve. However, this issue can be circumvented via
the minimization of the system's total energy by employing the
Bogomol'nyi-Prasad-Sommerfield (BPS) prescription \cite{n4, n4a}. The
implementation of {such} algorithm determines a specific expression for the
potential, {but it also provides} }a {lower bound {for the} energy (the BPS
bound) and the corresponding BPS equations whose solutions saturate that
bound (and therefore describe energetically stable configurations). In
addition, {it is always possible to verify that} the BPS equations are
compatible with the Euler-Lagrange equations, {from which one concludes that}
the BPS profiles stand for legitimate solutions {of} the model. In the
literature, there are alternative methods for the obtainment of such BPS
configurations; see, for instance, the study of the conservation of the
energy-momentum tensor \cite{ano}, the on-shell procedure \cite{onshell},
and the strong-necessary conditions technique \cite{lukasz, l1, l2, l3, l4}.}

{The} full Skyrme model was proposed in 1961 as a generalized nonlinear
sigma theory defined in $(3+1)$-dimensions {\cite{1a}}. Its Lagrange density
contains the so-called Skyrme term ({a quartic kinetic, i.e.} a term of
degree four in the first-derivative of the scalar sector), the $\sigma $
term (a quadratic kinetic one), and a potential which was originally proposed {as an attempt} to study the pion mass. {The Skyrme model {can be interpreted as} an effective low-energy model of Quantum Chromodynamics {which engenders} stable solitonic structures, so-called Skyrmions, {which} can be applied {to {the} study} some physical properties of those hadrons and nuclei {\cite{2a, 2a1, 2a2, 2a3, 2a4}}.} {Phenomenological applications of the gauged Skyrme model include not only the studies about the electromagnetic transition strengths for light nuclei \cite{x1} and the spin excitation energy of the nucleon \cite{x2}, but also investigations on the energy levels of a light nuclei $A=12$ \cite{x3}, the proton and neutron properties in a strong magnetic field \cite{x4} and, more recently, the electromagnetic transition rates of C$^{12}$ and O$^{16}$ in rotational-vibrational models \cite{x5}.}

In this context, the study of the planar version of the Skyrme theory, known
as the {baby Skyrme model \cite{3a}}, serves to the comprehension of many
aspects of the original $(3+1)$-dimensional scenario, including the
conditions under which it eventually accepts the implementation of the BPS
prescription. {The baby Skyrme model in the absence of the $\sigma $-term,
named the restricted baby Skyrme model \cite{13a}, supports a {%
well-established} BPS structure \cite{14a}.} Furthermore, over the last
years, the Skyrmions have also been used to describe topological quantum
Hall effect {\cite{4a, 4a1, 4a2, 4a3, 4a4}}, in chiral nematic liquid
crystals \cite{5a, 5a1}, superconductors {\cite{6a}}, brane cosmology {\cite%
{7a, 7a1, 7a2}} and magnetic materials {\cite{8a, 8a1}}, for instance.

Moreover, in order to investigate the electromagnetic properties of the baby
Skyrme model, it is necessary to couple it to {an Abelian gauge field \cite%
{16a}}. {In such a context, the BPS Skyrmions {\ appear in a} restricted
baby Skyrme{-Maxwell} model \cite{a1, a11, a12}, {and also occur} when {the
Skyrme sector is }minimally coupled to the Chern-Simons term \cite{a2} and
to the Maxwell-Chern-Simons action \cite{a3}. Additional results on the
study of those BPS solutions in a Skyrme-Born-Infeld scenario {can be found
in} \cite{a4}, while supersymmetric extensions of these restricted gauged
baby Skyrme {\ theories} are in the Refs. \cite{se, se1, se2, se21, se3,
se31, se4}.}

We now go a little bit further into this issue and consider how the
electromagnetic properties of a material medium affect the self-dual
Skyrmions which arise from a BPS restricted baby Maxwell-Skyrme model. Here,
these properties are studied via the introduction of a nonstandard function
which multiples the Maxwell term and therefore represents the magnetic
permeability of the medium.

{{To motivate} our study, we highlight that enlarged {models with a nontrivial} permeability have been considered with relative intensity in recent years. In the context of scalar field theories, for instance, it is currently known that the presence of such a permeability can be used to simulate geometrical constrictions in the corresponding kinklike solutions \cite{b22}, with the resulting profiles mimicking experimental {results and therefore clarifying} the influence of such a constriction on the magnetization in a magnetic material, see the Ref. \cite{b14}. Moreover, inspired by an experimental investigation on the possibility of controlling the domain wall polarity in a magnetic material in the presence of an electric pulse \cite{b15}, some authors have recently studied how the presence of geometric constrictions influences the behavior of fermions in a model with a nontrivial permeability, see the Ref. \cite{b23}.}

In order to present our results, this manuscript is organized as follows.
In the Section \ref{sec2}, we introduce the restricted baby Maxwell-Skyrme model enlarged via the inclusion of a nontrivial magnetic permeability. We
present the definitions and conventions which we adopt in our work. {In the sequence, we look for the BPS framework inherent to the
generalized scenario} via the minimization of the its total energy by means
of the BPS prescription. {As a result, it arises a differential
constraint (which we call \textit{superpotential equation}) which relates
the BPS potential to both the superpotential and the nontrivial
permeability. In view of such a constraint,} we obtain not only the BPS
bound {for the energy} itself, but also the self-dual equations
{whose solutions} saturate it. {We then particularize our work
by focusing} our attention on the gauged Skyrmions in a planar context,
{from which we rewrite the BPS equations in a rotationally symmetric
form. The Sec. \ref{sec3} is dedicated to the BPS scenario and its solutions. Here, in view of the target space inherent to a
Skyrme-Maxwell scenario, we fix an specific analytical expression for the
permeability which then forces the gauge sector to assume a nonusual shape.
In this context, we consider two different scenarios based on the ``nature"
of the superpotential, i.e. a first one in which the superpotential is given
by an exact expression, and a second case in which the superpotential must
be itself obtained numerically. In both cases, we work with potentials which
attain their vacuum values in the very same way, for the sake of comparison.
We then solve the two models numerically and depict the corresponding
profiles, wherefrom we identify how a noncanonical permeability may give
rise to BPS solutions with nonstandard shapes. We perform an analytical
study which explains the form which distinguishes the resulting magnetic
field.  We also consider some basic aspects of our generalized theory (in comparison to the standard case), such as the conditions under which BPS solutions do exist, whether the nontrivial model is capable to reproduce the BPS bound inherent to the ungauged baby Skyrme scenario and how a nontrivial permeability affects the value of the magnetic flux calculated for small and large electromagnetic coupling $g$. Next, we present some family of exact BPS solution for the enlarged model. Finally, the Sec. \ref{sec4} brings a brief summary and our perspectives regarding future contributions.}

In this manuscript, we adopt the {natural units system} and $\eta ^{\mu \nu
}=(+--)$ for the metric signature, for the sake of simplicity.


\section{The restricted gauged baby {Skyrme model in a} magnetic medium: The BPS
structure \label{sec2}}

\label{general}

We begin by presenting the $(2+1)$-dimensional restricted gauged baby Skyrme
model enlarged via the inclusion of {{an {a priori} arbitrary function which
represents a nontrivial} magnetic permeability, the corresponding Lagrangian
function reading%
\begin{equation}
L=E_{0}\int d^{2}\mathbf{x}\,{\mathcal{L}}\text{,}  \label{L0}
\end{equation}%
where the factor $E_{0}$ sets the energy scale {of the model} (which will be
taken {as} $E_{0}=1$ hereafter). The Lagrangian density is}
\begin{equation}
\mathcal{L}=-\frac{G}{4g^{2}}F_{\mu \nu }F^{\mu \nu }-\frac{\lambda ^{2}}{4}%
(D_{\mu }\vec{\varphi}\times D_{\nu }\vec{\varphi})^{2}-V\text{.}  \label{01}
\end{equation}
Here, the first term stands for Maxwell's action now multiplied by a magnetic permeability function {$G\equiv G(\varphi _{n})$ (this explicit dependence on the quantity $\varphi _{n}=\hat{ n } \cdot \vec{\varphi}$ will be clarified later during the implementation of the BPS formalism).} In the internal space, $\hat{n}$ represents an unitary vector which defines a preferred direction, while the Skyrme field $\vec{\varphi}=(\varphi_{1},\varphi _{2},\varphi _{3})$ is given as a triplet of real scalar fields constrained to satisfy $\vec{\varphi}\cdot \vec{\varphi}=1$ and therefore describing a spherical surface with unitary radius. Moreover, $F_{\mu \nu }=\partial _{\mu }A_{\nu }-\partial _{\nu }A_{\mu }$ is the electromagnetic field strength tensor and%
\begin{equation}
D_{\mu }\vec{\varphi}=\partial _{\mu }\vec{\varphi}+A_{\mu }\hat{n}%
\times \vec{\varphi}
\end{equation}%
stands for the usual covariant derivative of the Skyrme field. {The third term brings the self-interacting potential $V=V(\varphi _{n})$ which promotes the spontaneous breaking of the internal symmetry. At the same time, }both $\lambda $ and $g$ are coupling constants inherent to the model (which we assume to be nonnegative from now on). Moreover, the Skyrme field and the function $G$ are dimensionless,  while the gauge field, the electromagnetic constant $g$ and the Skyrme one $\lambda$ have mass dimensions equal to $1$, $1$, and $-1$, respectively.

{{Now, beyond the} motivations previously cited in the Section \ref{sec1}, we also point out that the idea based on the inclusion of a nontrivial permeability has also been widely used in the context of gauged models, with different purposes: for instance, {the Refs. \cite{b1, b2} applied it to describe a bag model} similar to the MIT \cite{b3} and SLAC \cite{b4} bag models, while some authors have implemented the same idea to study peculiar properties of gauged vortices; see the Refs. \cite{b5, b6,b7}. In addition, in the Refs. \cite{b8, b9, b10, b11}, a nontrivial permeability was used in connection with the AdS/CFT correspondence.} {More recently, it was also employed to study both the presence of electrically charged structures in a multi-field scenario \cite{b27}, and the arising of internal structures in Abelian gauge field models generated by both an electric point charge \cite{b16} and an electric dipole \cite{bprd} when immersed in a medium controlled by scalar fields. In particular, recent studies \cite{b12, b13, b14a, b15a} on dielectric Skyrme models were investigated in view of their possible connections with the binding energies of nuclei.}

{{Here, it is worthwhile to} clarify that the Lagrange density (\ref{01}) must not be considered as a trivial generalization of the model investigated in the Ref. \cite{a1} once that, as we demonstrate below, {the magnetic permeability composes the differential constraint involving both the self-dual potential and the superpotential, which provides support for the existence of the BPS structure. Consequently,} it is possible to modify the vacuum structure of the effective model {by conveniently choosing} the form of the magnetic permeability, from which {configurations with different shapes and features may occur, for instance.}}

{It is instructive to write down the Gauss law for time-independent
configurations which comes from (\ref{01}), i.e}%
\begin{equation}
\partial _{i}\left( G\partial ^{i}A^{0}\right) =-g^{2}\lambda ^{2}A_{0} \left( \hat{n}\cdot \partial_{i}\vec{\varphi}\right)^2.
\label{gl}
\end{equation}%
The point here is that $A^{0}=0$ stands for a legitimate gauge choice, given that it solves the Gauss law (\ref{gl}) identically. Thus, we conclude that the stationary configurations we study in this manuscript are electrically neutral (i.e. present no electric field and electric charge).

Instead of studying the solutions of the second-order Euler-Lagrange equations, we focus our attention on those BPS configurations that minimize the theory's total energy. Here, we achieve such a goal via the implementation of the BPS procedure whose starting point is the stationary energy density of the model (\ref{01}),
\begin{equation}
\varepsilon =\frac{G}{2g^{2}}B^{2}+\frac{\lambda ^{2}}{2}Q^{2}+V%
\text{,}  \label{ed00}
\end{equation}%
{where we have already implemented $A_{0}=0$ and $Q$ defined by
\begin{eqnarray}
Q&=&\vec{\phi}\cdot (D_{1}\vec{\phi}\times D_{2}\vec{\phi})\notag\\[0.2cm]
&=&\vec{\varphi}\cdot (\partial _{1}\vec{\varphi}\times\partial_{2} \vec{\varphi})+\epsilon_{ij}A_{i}(\hat{n}\cdot \partial _{j}\vec{\phi}). \label{QQ}
\end{eqnarray}
where the term $\vec{\varphi}\cdot (\partial_{1}\vec{\varphi}\times \partial _{2}\vec{\varphi})$ is related to the topological charge of the Skyrme field by means of
\begin{equation}
\deg \left[ \vec{\varphi}\right] =-\frac{1}{4\pi }\int d^{2}\mathbf{x~}\vec{%
\varphi}\cdot (\partial _{1}\vec{\varphi}\times \partial _{2}\vec{\varphi})=k
\text{,}
\end{equation}%
where $k\in \mathbb{Z\setminus }\left\{ 0\right\}$.}

{We now establish the boundary conditions to be satisfied by the fields to ensure the existence of finite energy configurations. For this, the energy density (\ref{ed00}) must be zero at the vacuum, i.e. when  $|\mathbf{x}| \rightarrow \infty$. Consequently, the magnetic field $B$, $Q$ and the potential $V$ must satisfy the following boundary conditions:
\begin{equation}
\lim_{|\mathbf{x}|\rightarrow\infty}\sqrt{G}B=0, \;\; \lim_{|\mathbf{x}| \rightarrow\infty} Q=0,\;\; \text{and}\;\;\lim_{|\mathbf{x}|\rightarrow\infty} V=0 \text{.}  \label{bcBQU}
\end{equation}}

The total energy $E$ is defined as the integral of the energy density (\ref{ed00}), so that the implementation of the BPS formalism allows us to
write
\begin{eqnarray}
E &=&\int d^{2}\mathbf{x}\left[ \frac{\left(GB\pm\lambda^{2}g^{2}\mathcal{W} \right)^{2}}{2Gg^{2}}+\frac{\lambda ^{2}}{2}\left(Q\pm \frac{\partial \mathcal{W}}{\partial \varphi_{n}}\right)^{2} \right.  \notag \\[0.2cm]
&&\hspace{1cm}\mp \lambda ^{2}B\mathcal{W}\mp \lambda ^{2}Q\frac{\partial \mathcal{W}}{\partial \varphi_{n}} \notag \\[0.2cm]
&&\hspace{1cm}\left. -\frac{\lambda ^{4}g^{2}}{2G}\mathcal{W}^{2} -\frac{\lambda ^{2}}{2}\left(\frac{\partial \mathcal{W}}{\partial \varphi_{n}} \right)^{2}+V\right] \text{,}
\quad  \label{en1}
\end{eqnarray}%
where we have introduced {the superpotential function  $\mathcal{W}\equiv \mathcal{W}(\varphi _{n})$.  Moreover, it proves useful to define the BPS
potential $V(\varphi_{n})$ as}
\begin{equation}
V=\frac{\lambda ^{4}g^{2}}{2G}\mathcal{W}^{2}+\frac{\lambda ^{2}}{2}\left( {%
\frac{\partial \mathcal{W}}{\partial \varphi _{n}}}\right) ^{2}\text{,}
\label{Vv}
\end{equation}%
from that one gets that the third row of the Eq. (\ref{en1})
vanishes. {{It is worthwhile to note that the Eq. (\ref{Vv}) also explains why $G$ is supposed to depend on  $\varphi _{n}$ only, i.e., given the superpotential $\mathcal{W}=\mathcal{W}(\varphi_{n})$ and to maintain the dependence of the potential ${V}$  on $\varphi _{n}$ (which stands for a necessary condition to the formation of gauged Skyrmions), we must choose $ G=G(\varphi_{n})$.}} Furthermore, when the Skyrme field assumes its vacuum configuration (supposed to be $\varphi_{n} \rightarrow 1$, as usual), the potential vanishes, and the Eq. (\ref{Vv})  leads to
\begin{equation}
\lim_{\varphi _{n}\rightarrow 1}{\mathcal{W}}{(\varphi _{n})}=0\text{ \ \
and \ \ }\lim_{\varphi _{n}\rightarrow 1}\frac{\partial \mathcal{W}}{%
\partial \varphi _{n}}=0\text{,}  \label{BcVv}
\end{equation}%
{i.e. the boundary conditions to be satisfied for the superpotential itself.}
{After using the explicit form of $Q$ given by Eq. (\ref{QQ}),  the total energy  (\ref{en1}) assumes the form}
\begin{eqnarray}
E &=&\int d^{2}\mathbf{x}\left[ \frac{\left( GB\pm \lambda ^{2}g^{2}\mathcal{%
W}\right) ^{2}}{2Gg^{2}}+\frac{\lambda ^{2}}{2}\left( Q\pm \frac{\partial \mathcal{W}}{\partial \varphi_{n}}\right)^{2}\right.  \notag \\[0.2cm]
&&\hspace{0cm}\left.\frac{}{}\mp \lambda ^{2}\frac{\partial \mathcal{W}}{\partial \varphi_{n}}\vec{\varphi}\cdot (\partial _{1}\vec{\varphi}\times\partial _{2}\vec{\varphi})\mp \lambda ^{2}\epsilon _{ij}\partial _{j}(\mathcal{W}A_{i}) \right]\!\! .\quad  \label{enNN}
\end{eqnarray}

{The point is that, given the boundary conditions (\ref{BcVv}), the contribution due to the total derivative $\epsilon _{ij}\partial _{j}(\mathcal{W}A_{i})$ appearing in  Eq. (\ref{enNN}) vanishes. In this way, we can express the total energy as,}
\begin{equation}
E=\bar{E}+E_{bps}\text{,}  \label{en5}
\end{equation}%
where $\bar{E}$ represents {the integral composed} by the quadratic terms, {%
i.e.}%
\begin{equation}
\bar{E}=\int d^{2}\mathbf{x}\left[ \frac{\left( GB\pm \lambda ^{2}g^{2}%
\mathcal{W}\right) ^{2}}{2Gg^{2}}+\frac{\lambda ^{2}}{2}\left( Q\pm \frac{%
\partial \mathcal{W}}{\partial \varphi _{n}}\right) ^{2}\right]\! \text{,}\quad
\label{en4}
\end{equation}%
and $E_{bps}$ defines the energy lower bound, {which reads}%
\begin{equation}
E_{bps}=\mp \lambda ^{2}\int d^{2}\mathbf{x}\left( \frac{\partial \mathcal{W}%
}{\partial \varphi _{n}}\right) \vec{\varphi}\cdot (\partial _{1}\vec{\varphi%
}\times \partial _{2}\vec{\varphi})>0\text{.}  \label{en3}
\end{equation}%

{Given that} $\bar{E}\geq 0$, the total energy (\ref{en5}) {satisfies}
the {typical BPS} inequality%
\begin{equation}
E\geq E_{bps}\text{,}
\end{equation}%
{from which we conclude that} {the energy lower bound {is }achieved
when the fields {are} such that $\bar{E}=0$, i.e. when }{they
satisfy}%
\begin{eqnarray}
GB &=&\mp g^{2}\lambda ^{2}\mathcal{W}\text{,}  \label{be10} \\[0.2cm]
Q &=&\mp \frac{\partial \mathcal{W}}{\partial \varphi _{n}}\text{,}
\label{be20}
\end{eqnarray}%
{which therefore stand for the self-dual equations inherent to the enlarged model.} {The solutions of these equations also are classical solutions belonging to an extended supersymmetric model \cite{witten, spector} whose bosonic sector would be given by the Lagrangian density (\ref{01}).} {Furthermore, some studies concerning the gauged Skyrme model in the SUSY field theory context can be found, for instance, in Refs. \cite{41w, 42w, 43w, 44w}.}

{The interested reader {must note that,} beyond multiplying the magnetic field  $B$ in the self-dual Eq. (\ref{be10}) {(an expected fact given the way how the magnetic permeability appears in the Lagrangian density (\ref{01})), the function $G$ also composes the Eq. (\ref{Vv}) relating the BPS potential $V(\varphi_{n})$ and the superpotential $\mathcal{W}(\varphi_{n})$.}}

{{Moreover, the combination of Eqs. (\ref{ed00}) and (\ref{Vv}) together with the self-dual ones (\ref{be10}) and (\ref{be20})} leads to the following expression for the BPS energy density:} {
\begin{equation}
\varepsilon _{bps}=\frac{G}{g^{2}}B^{2}+\lambda ^{2}Q^{2}\text{,}
\label{ebps1a}
\end{equation}
which shows clearly the contribution of the magnetic permeability.}

{\subsection{Rotationally symmetric BPS Skyrmions}}

{Once we have developed the general BPS framework, we focus our investigation on those solutions with rotational symmetry. In this sense,} without loss of generality, we set $\hat{n}=(0,0,1)$, wherefrom we get $\varphi_{n}=\varphi _{3}$. As a consequence, the potential $V=V(\varphi_{3})$ now allows for the spontaneous breaking of the $SO(3)$ symmetry inherent to the Skyrme-Maxwell model (\ref{01}) that enables the occurrence of configurations with a nontrivial topology as expected.

Moreover, in order to compare our results with the well-established ones, we
study time-independent solutions using the standard ansatz for the gauge field
\begin{equation}
A_{i}=-\epsilon _{ij}\hat{x}_{j}\frac{Na\left( r\right) }{r}\text{,}
\end{equation}
and the Skyrme field
\begin{equation}
\vec{\varphi}=\left(
\begin{array}{c}
\sin f\cos \left( N\theta \right) \\
\sin f\sin \left( N\theta \right) \\
\cos f%
\end{array}%
\right) \text{,}
\end{equation}%
where $r$ and $\theta $ are polar coordinates, $\epsilon _{ij}$ stands for
the Levi-Civita antisymmetric tensor (with $\epsilon _{12}=+1$) and $\hat{x}_{i}=(\cos \theta ,\sin \theta )$ represents the unit vector. Also, $N$ is the winding number of the Skyrme field, while the profile functions $f(r)$
and $a(r)$ are supposed to obey the boundary conditions which are known to
support the existence of regular solutions with finite energy,
\begin{eqnarray}
f\left( r=0\right) =\pi \text{ \ } &\text{and}&\text{ \ }f\left(
r\rightarrow \infty \right) \rightarrow 0\text{,}  \label{bc1} \\[0.2cm]
a\left( r=0\right) =0 &\text{ \ and}&\text{ \ }a^{\prime }\left(
r\rightarrow \infty \right) \rightarrow 0\text{,}  \label{bc2}
\end{eqnarray}%
in which prime denotes the derivative with respect to the radial coordinate $%
r$.

{It is instructive to point out that the} magnetic field in terms of
the ansatz reads
\begin{equation}
B(r)=F_{21}=-\frac{N}{r}\frac{da}{dr}\text{.}  \label{mf}
\end{equation}

{In what follows, for the sake of} convenience, {we} implement
the field redefinition%
\begin{equation}
h(r)=\frac{1}{2}\left( 1-\cos f\right) \text{,}  \label{rf}
\end{equation}%
from where one gets that the new profile function $h(r)$ satisfies the boundary conditions%
\begin{equation}
h\left( r=0\right) =1  \text{ \ and} \text{ \ }h\left( r\rightarrow \infty \right)
\rightarrow 0\text{.}  \label{bc3}
\end{equation}

{In view of the Eq. (\ref{rf}),} {both} $G$ and $\mathcal{W}$ become {%
functions} of $h$ only. {In particular,} the boundary conditions
{to be} satisfied by ${\mathcal{W}}{(h)}$ {can be summarized as}%
\begin{equation}
\lim_{r\rightarrow 0}{\mathcal{W}}{(h)}={\mathcal{W}}_{0}\text{,}\ \
\lim_{r\rightarrow \infty }{\mathcal{W}}{(h)}=0\text{,}\ \
\lim_{r\rightarrow \infty }\frac{\partial \mathcal{W}}{\partial h}=0\text{,}
\label{bbccW}
\end{equation}%
where $\mathcal{W}_{0}>0$, whereas the two last ones  correspond to those which appear in {Eq. (\ref{BcVv})}.

The BPS energy {given by} {Eq. (\ref{en3})} {can be calculated explicitly,
its value reading}%
\begin{equation}
E_{bps}=\mp 2\pi \lambda ^{2}N{\mathcal{W}}_{0}>0\text{,}  \label{en3x}
\end{equation}%
Here, the upper (lower) sign corresponds to $N<0$ ($N>0$).

The BPS equations (\ref{be10}) and (\ref{be20}) become%
\begin{equation}
B=-\frac{N}{r}\frac{da}{dr}=\mp \frac{\lambda ^{2}g^{2}\mathcal{W}}{G}  \label{be1}
\end{equation}
\begin{equation}
\frac{\left( 1+a\right) }{r}\frac{dh}{dr}=\pm \frac{1}{4N}\frac{\partial
\mathcal{W}}{\partial h}\text{,}  \label{be2}
\end{equation}%
{respectively, where we have used the Eq. (\ref{mf}) for the magnetic field.}

To summarize, these equations {above} describe a radially symmetric
structure whose total energy is {given by Eq. (\ref{en3x})}. {Further,} the
gauged Skyrmions emerge as the numerical solutions of the BPS equations (\ref{be1}) and (\ref{be2}) obtained via the boundary conditions (%
\ref{bc2}) and (\ref{bc3}).

In the next Sections, we demonstrate how the BPS framework introduced
{here} can be used to generate legitimate gauged Skyrmions in the
presence of a nontrivial {magnetic permeability}. {Additionally, we
also investigate some basic properties of our enlarged model, in comparison
to those presented by its canonical version.}

\vspace{0.75cm}

\section{BPS Skyrmions in magnetic media\label{sec3}}

{We now particularize our investigation by focusing our attention on some effective models. {Therefore, the results presented below can contribute to the understanding of the electromagnetic properties of gauged Skyrmions by studying their BPS-gauged} versions. The point here is that exploring the electromagnetic properties of the Skyrmions is commonly a rather complicated work {even in a non-BPS context, as shown by some currently available results, see, for instance, the} Refs. \cite{48w, 49w} for correlated developments within the standard gauged Skyrme model.} {{In particular, it becomes clear that the interaction with an Abelian gauge field plays a fundamental role concerning the properties of baryons and atomic nuclei. In that regard, it is known, for instance, that while the exact form of the {low-energy Skyrme} theory remains unknown, its coupling to the electromagnetic sector is already fixed, see the Refs. \cite{45w, 46w, 47w}.}}

{Concerning the (1+2)-dimensional case, Refs. \cite{39w,39wb} have studied some aspects of the magnetic properties arising in gauged BPS baby Skyrmions. This way, the enlarged scenario plans to provide new results about BPS baby Skyrmions immersed in a magnetic medium. As we have already argued, we intend to identify the new effects produced on BPS Skyrmions due to a nontrivial magnetic permeability. In particular, we look for the arising of Skyrmions with internal structures. With such aimin mind, we choose the permeability as}
\begin{equation}
G(h)=\frac{1}{\left( \gamma -h^{2}\right) ^{\beta }}\text{.}  \label{gh0}
\end{equation}%
{Here, $\gamma, \beta\in \mathds{R}$, with $\beta \geq 0$. Thus, the} rotationally symmetric version of the BPS equations (\ref{be1}) and (\ref{be2}) can be written in the form
\begin{equation}
N\frac{da}{dy}=\pm \lambda ^{2}g^{2}\left( \gamma -h^{2}\right) ^{\beta }%
\mathcal{W}\text{,}  \label{bpsx1}
\end{equation}%
\begin{equation}
\left( 1+a\right) \frac{dh}{dy}=\pm \frac{1}{4N}\frac{d\mathcal{W}}{dh}\text{%
,}  \label{bpsx2}
\end{equation}%
{where we have introduced a new spatial coordinate $y$ defined by $y=r^{2}/2$.}

\begin{figure}[tbp]
\includegraphics[width=8.4cm]{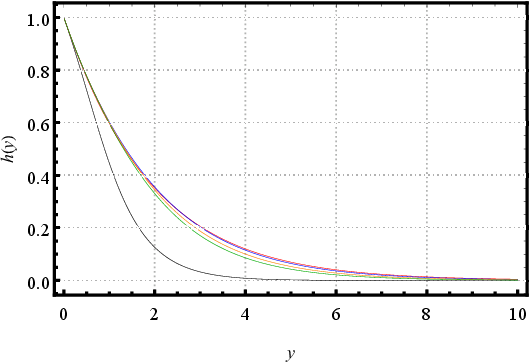}\vspace{0.5cm} %
\includegraphics[width=8.4cm]{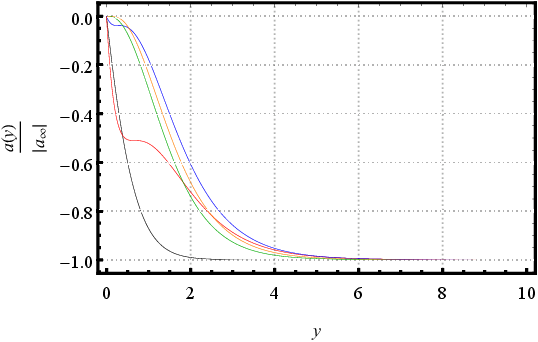}
\caption{Numerical solutions to both $h(y)$ (top) {and }$a(y)$ (bottom,
depicted in units of $\left\vert a_{\infty }\right\vert $) obtained from the
Eqs. (\ref{cx01x}) and (\ref{cx02x}) for $\gamma %
=0.50$\ (red line), $\gamma =0.75$ (blue line), $\gamma =0.90
$ (orange line) and $\gamma =1.00$ (green line). The corresponding
usual profile (obtained via $G=1$) appears as the black line, for the sake
of comparison.}
\end{figure}

{In what follows, we split our investigation into two different branches according to the nature of the superpotential $\mathcal{W}(h)$. Within this sense, in Sec. \ref{sec3A}, we first consider a particular scenario for which the superpotential is given {by an explicit function of $h(r)$, so that the Eq. (\ref{Vv}) provides the BPS potential in terms of $\mathcal{W}(h)$. Next, we use it to solve the corresponding BPS equations numerically. As a second case, in Sec. \ref{sec3B}, we select explicitly the functional form of $V(h)$, so that the Eq.  (\ref{Vv})  becomes a differential equation for the superpotential $\mathcal{W}(h)$, which, in general, must be solved numerically together with the BPS equations.} We then present some considerations about comparing the resulting profiles emerging from these two contexts.}

{\subsection{Analytical superpotential and its BPS solutions\label{sec3A}}}

{In order to solve the BPS equations, we first consider the superpotential as}
\begin{equation}
\mathcal{W}(h)=\frac{h^{2}}{\lambda ^{2}}\text{,}  \label{w10}
\end{equation}%
from which one gets $\mathcal{W}_{0}=\lambda ^{-2}>0$ and the total energy of the BPS configurations as $E_{bps}=2\pi \left\vert N\right\vert >0$, as expected, see the Eq. (\ref{en3x}). It is also clear that the superpotential above satisfies the conditions given in the Eq. (\ref{bbccW}). This choice was motivated by the fact that the superpotential (\ref{w10}) is known to support  well-behaved Skyrmions which attain their asymptotic values  according to  a Gaussian decay law, as explained recently in the Refs. \cite{a2,a3,a4}.

{It is instructive to consider the potential $V(h)$ related to $\mathcal{W}(h)=\lambda ^{-2}h^{2}$. With this aim in mind, we write the superpotential equation (\ref{Vv}) as}
\begin{equation}
\mu ^{2}U(h)=\frac{\lambda ^{4}g^{2}}{2G}\mathcal{W}^{2}+\frac{\lambda ^{2}}{%
8}\left( {\frac{d\mathcal{W}}{dh}}\right) ^{2}\text{,}  \label{vvrs}
\end{equation}
{where we have rescaled the potential $V(h)$ as $\mu^{2}U(h)$, for the sake of comparison between our results and the standard ones. Solving the Equation above for $U(h)$, we obtain}
\begin{equation}
U(h)=\frac{h^{2}}{2\mu ^{2}\lambda ^{2}}\left[ 1+\lambda ^{2}g^{2}\left(
\gamma -h^{2}\right) ^{\beta }h^{2}\right] \text{,}  \label{ap}
\end{equation}%
{where we have also considered the Eq. (\ref{gh0}) for $G(h)$.}

It is interesting to note that, in the limit $h\left( r\rightarrow \infty
\right) \rightarrow 0$, the generalized potential above approaches the
vacuum as%
\begin{equation}
U(h\rightarrow 0)\approx \frac{h^{2}}{2\mu ^{2}\lambda ^{2}}\text{,}
\label{uh0}
\end{equation}%
i.e. in the very same way as its standard counterpartner.\ As a consequence,
we conclude that, despite the nontrivial expression which we have chosen for
the magnetic permeability, a superpotential of the form $\mathcal{W}%
(h)\propto h^{2}$ leads to a potential which behaves as $U(h)\propto h^{2}$
in the asymptotic region, and vice-versa (we return to such a conclusion
later below).

Now, in view of the Eq. (\ref{w10}), the BPS equations {(\ref{bpsx1})
and (\ref{bpsx2}) assume the form}%
\begin{equation}
N\frac{da}{dy}=\pm g^{2}\left( \gamma -h^{2}\right) ^{\beta }h^{2}\text{,}
\label{bpsx11}
\end{equation}%
\begin{equation}
\left( 1+a\right) \frac{dh}{dy}=\pm \frac{h}{2\lambda ^{2}N}\text{,}
\label{bpsx22}
\end{equation}%
via which we intend to investigate those gauged {Skyrmions} {which} behave
standardly at the boundaries and have a noncanonical {profile} for
intermediate values of $y$. As we clarify below, this type of configuration
is directly related to different values of $\gamma $, from which we work
with fixed values for the others parameters. In particular, we {{set} $\beta
=2$}, $g=1$, $\lambda =1$ and $N=1$ (i.e. the lower signs in the BPS
equations), for {the sake of} simplicity.

\begin{figure}[tbp]
\includegraphics[width=8.4cm]{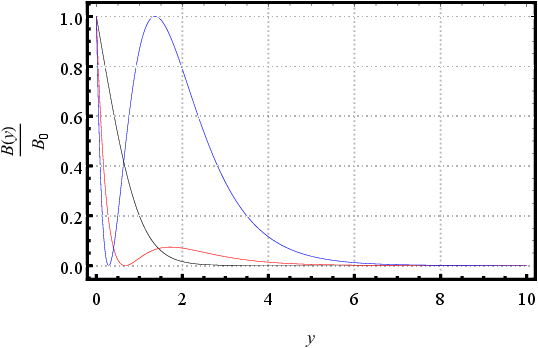}\vspace{0.5cm} %
\includegraphics[width=8.4cm]{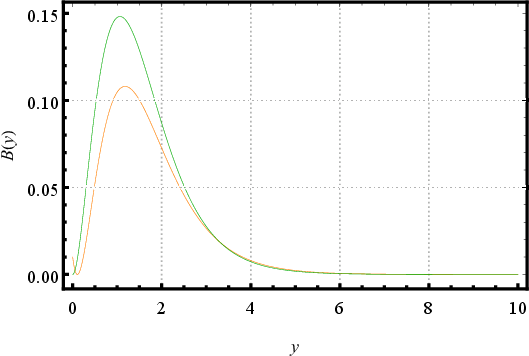}
\caption{Numerical solutions to the BPS magnetic field $B(y)$ obtained from
the Eqs. (\ref{cx01x}) and (\ref{cx02x}). Conventions as in
the Fig. 1. Top: the profiles (depicted in units of $B_{0}=B(y=0)$) for $%
\gamma =0.50$ and $\gamma =0.75$. Bottom: the results for $%
\gamma =0.90$ and $\gamma =1.00$. The corresponding standard
solution again appears as the black line. Here, some of the noncanonical
profiles were normalized for the sake of visualization.}
\end{figure}

In view of these choices, the BPS Eqs. (\ref{bpsx11}) and (\ref{bpsx22})
reduce to%
\begin{equation}
\frac{da}{dy}=-h^{2}\left( \gamma -h^{2}\right) ^{2}\text{,}  \label{cx01x}
\end{equation}%
\begin{equation}
\left( 1+a\right) \frac{dh}{dy}=-\frac{1}{2}h\text{,}  \label{cx02x}
\end{equation}%
which must be solved numerically via the implementation of a
finite-difference scheme together {with} the boundary conditions (\ref{bc2})
and (\ref{bc3}).

The Figure {1 shows the numerical solutions to both }$h(y)$ (top) {and} $%
a(y) $ (bottom) {for different values of} $\gamma $. {Here, the gauge
profile function }$a(y)${\ is depicted in units of }$\left\vert
a_{\infty }\right\vert ${, with }$a_{\infty }=a(y\rightarrow \infty )$%
{.}

It is now clear how $\gamma $ affects the size of the core of $h(y)$ in an
inverse way, i.e. as the values of $\gamma $ increase, the resulting core
decreases.\ In\ addition, regarding the gauge profile function, it is
worthwhile to note that the solutions with $\gamma <1$ are characterized by
the presence of noncanonical plateaus which appear for intermediate values
of the coordinate $y$. In this sense, we point that our numerical
investigation has revealed that, for $\gamma >3$, both $h(y)$ and $a(y)$
tend to compactify as $\gamma$ increases. Such a behavior is analogue
to that already found  in the standard case ($G=1$) for increasing values of
the coupling constant $g$.

In the Fig. 2, we show the numerical solutions to the BPS\ magnetic field $%
B(y)$, from which it is possible to see how the shape of this field depends
on the value of $\gamma $ in a dramatic way. In the sequence, we proceed
with an analytical study of such a dependence, via which we clarify how the
aforecited plateaus give rise to the formation of {nonstandard} internal
structures which distinguish the behavior of the corresponding magnetic
sector.

In order to study the way $\gamma $ affects the shape of $B(y)$, we write
this field as%
\begin{equation}
B(y)=h^{2}\left( \gamma -h^{2}\right) ^{2}\text{,}  \label{be021}
\end{equation}%
whose first derivative provides%
\begin{equation}
\frac{dB}{dy}=2h\left( \gamma -h^{2}\right) \left( \gamma -3h^{2}\right)
\frac{dh}{dy}\text{.}
\end{equation}

Now, once the solution to the Skyrme profile function $h(y)$ is supposed to
vary monotonically from $1$ (at $y=0$) to $0$ (in the limit $y\rightarrow
\infty $, i.e. $h_{y}$ is always negative), {one gets that} the condition $%
B^{\prime }(Y)=0$ provides the extreme points (note that we are looking for
intermediary values of $y$, from which we are here excluding both the origin
and the asymptotic limit)%
\begin{equation}
h(Y_{1})=h_{1}=\sqrt{\gamma }<1\text{,}  \label{h1x}
\end{equation}%
\begin{equation}
h(Y_{2})=h_{2}=\sqrt{\frac{\gamma }{3}}<1\text{,}  \label{h2x}
\end{equation}%
where $0<Y_{1}<Y_{2}$.

{At these points,} the magnetic field assumes the values%
\begin{equation}
B_{1}=B\left( h_{1}\right) =0\text{,}  \label{bb1x}
\end{equation}%
\begin{equation}
B_{2}=B\left( h_{2}\right) =\frac{4}{27}\gamma ^{3}\text{,}  \label{bb2x}
\end{equation}%
respectively.

The first value, $B_{1}$, becomes a local minimum if $\gamma <1$, whereas $%
B_{2}$ results {in} a local maximum if $\gamma <3$. {Moreover,} from {Eq. (%
\ref{be021})}, the value of the magnetic field at the origin {is given by}%
\begin{equation}
B_{0}=B\left( y=0\right) =\left( \gamma -1\right) ^{2}\text{.}  \label{bb3x}
\end{equation}

In what follows, we use the Eqs. (\ref{bb1x}), (\ref{bb2x}) and (\ref{bb3x})
above to enumerate three different pictures based on the values of $\gamma $%
. The interested reader can apply the same prescription in order to describe
additional configurations with different $\gamma $. Here, it is important to
emphasize that we are considering intermediate values of $y$, i.e. we are
excluding $y=0$ and those values located in the {asymptotic} region $%
y\rightarrow \infty $.

\begin{figure}[tbp]
\includegraphics[width=8.4cm]{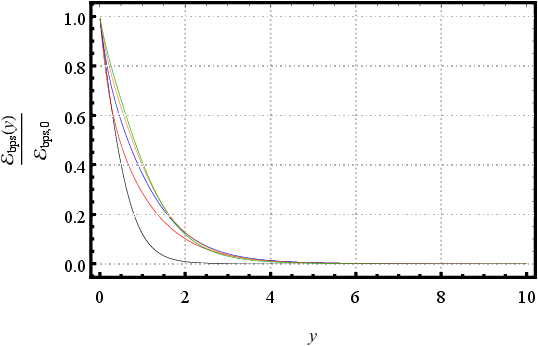}\vspace{0.5cm}
\caption{Numerical solutions to the energy density of the BPS
configurations, i.e. $\varepsilon _{bps}(y)$ (depicted in units of $%
\varepsilon _{bps,0}=\varepsilon _{bps}(y=0)$). Conventions
as in the Fig. 1. The profiles were normalized for the sake of
visualization. }
\end{figure}

\vspace{0.5cm}

{\subsubsection{$\gamma=0$ case} }

The first picture is defined for $\gamma =0$, from which one gets that the
BPS equations (\ref{cx01x}) and (\ref{cx02x}) assume the form%
\begin{equation}
\frac{1}{r}\frac{da}{dr}=-h^{6}\text{,}
\end{equation}%
\begin{equation}
\frac{\left( 1+a\right) }{r}\frac{dh}{dr}=-\frac{1}{2}h\text{,}
\end{equation}%
which, in view of the redefinition $H(r)=\left[ h(r)\right] ^{3}$, can be
written as%
\begin{equation}
\frac{1}{r}\frac{da}{dr}=-H^{2}\text{,}  \label{bx01x}
\end{equation}%
\begin{equation}
\frac{\left( 1+a\right) }{r}\frac{dH}{dr}=-\frac{3}{2}H\text{.}
\label{bx02x}
\end{equation}

In this case, despite the redefinition applied on the Skyrme profile
function, we note that the resulting Eqs.  (\ref{bx01x}) and (%
\ref{bx02x}) can be obtained directly from the general ones (\ref{be1}) and (%
\ref{be2}) for $G=1$, $g=N=1$ and $\lambda =\sqrt{1/3}$. We then conclude
that the a priori nontrivial case defined by $G(h)=h^{-4}$ stands for a
merely redefinition of the usual case (defined by $G=1$) with a different
value of the coupling constant $\lambda $. As a consequence, we do not
expect significant changes to occur on the shape of the solutions,
especially on that of the magnetic sector. Therefore, in what follows, we
consider only the case with nonvanishing values of $\gamma $.

\vspace{0.5cm}

{\subsubsection{$0<\gamma <1$ case} }

A second picture occurs when $0<\gamma <1$. In this context, the solution (%
\ref{h1x}) is satisfied at some point $y=Y_{1}$. At this point, the magnetic
field vanishes (i.e. $B\left( y=Y_{1}\right) =0$, see the {Eq.} (\ref{bb1x}%
)), from which it is reasonable to infer that the magnetic solution
describes {a centered lump surrounded by a ring}: the {lump} is {positioned}
at the origin, its amplitude being given by the {Eq.} (\ref{bb3x}) itself,
while the {radius of the ring} is {located }at some point $y=Y_{2}>Y_{1}$
(defined according to the {Eq.} (\ref{h2x})), {the amplitude of the ring}
standing for $B\left( y=Y_{2}\right) =\left( 4/27\right) \gamma ^{3}$, see
the previous Eq. (\ref{bb2x}).

We highlight how $\gamma $ determines the difference between these two {%
amplitudes}:\ for $0<\gamma <0.75$, {the magnitude of the centered lump} is
taller than {that} {of the ring} (i.e. $\left( \gamma -1\right) ^{2}>\left(
4/27\right) \gamma ^{3}$). On the other hand, when $\gamma =0.75$, the two {%
magnitudes} reach the very same value. Finally, for $0.75<\gamma <1$, the
amplitude {of the ring }is taller than {that of the lump positioned} at $y=0$
(i.e. $\left( \gamma -1\right) ^{2}<\left( 4/27\right) \gamma ^{3}$).

The parameter $\gamma $ also controls the values of both $Y_{1}$ (i.e. the
point at which $B(y)$ vanishes) and $Y_{2}$ (the {radius of the ring}): as $%
\gamma $ increases, the values of $h_{1}=\sqrt{\gamma }$ and $h_{2}=\sqrt{%
\gamma /3}$ also increase and, once $h(y)$ varies monotonically from $1$ to $0$, both $Y_{1}$ and $Y_{2}$ decrease (i.e. move toward the origin).

\vspace{0.5cm}

{\subsubsection{$\gamma =1$ case}}

The case with $\gamma =1$ defines another picture, for which {Eq. (\ref{h1x})%
} holds at the origin only (i.e. $h\left( y=0\right) =h_{1}=1$), which
agrees with the boundary condition (\ref{bc3}). Therefore, the magnetic
field vanishes at $y=0$, which agrees with the result which comes from the {%
Eq. (\ref{bb3x})} for $\gamma =1$. We then conclude that the resulting
magnetic profile stands for a {single ring} {whose radius} is located at
some point $y=Y_{2}$ (defined by $h\left( y=Y_{2}\right) =h_{2}=\sqrt{1/3}$,
see the Eq. (\ref{h2x})), its magnitude being equal to $B\left(
y=Y_{2}\right) =4/27$, see the Eq. (\ref{bb2x}).

As we said before, the very same prescription can be used by the reader to
describe additional configurations with different values of $\gamma $.

{The following section explores a more complex context where the superpotential is numerically determined. To achieve this goal, we consider the BPS and superpotential equations parts of a self-dual system, which we solve using numerical methods. Afterward, we compare the analytical and numerical approaches by commenting on the main characteristics.}

\vspace{0.5cm}

{\subsection{Numerical superpotential and its BPS solutions\label{sec3B}}}

{Following the idea introduced in \cite{a1}, we transform the Eq. (\ref{Vv}) in a differential equation for the superpotential $\mathcal{W}(h)$ which, together with the BPS equations (\ref{be1}) and (\ref{be2}), forms a set of differential equations to be solved for specific choices of both $V$ and $G$. Additionally, the boundary conditions satisfied by the profiles $a(r)$, $h(r)$, and $\mathcal{W}(h)$} {remain unaltered, from what it is possible to say that introducing a magnetic permeability does not change the target space of the effective model, as expected.}

\begin{table}[t]
\caption{Approximate numerical values used for the normalization of the noncanonical
profiles obtained from the Eqs. (\ref{cx01x}) and (\ref%
{cx02x}). The standard values (with $%
\beta =0$) are $a_{\infty }=-0.632121$, $B_{0}=1$ and $\varepsilon _{bps,0}=2$, for the sake of comparison.}%
\begin{ruledtabular}
		\begin{tabular}{cccc}
	$\gamma$&$\quad a_\infty$&$B_{0}$&$\varepsilon _{bps,0}$\\
			\colrule
			&  & & \\ [-0.25cm]	
			{0.50}&$-0.079951$&$0.2500$&{1.2500} \\
			{0.75}&$-0.135692$&$0.0625$&{1.0625} \\
			{0.90}&$-0.215987$&$0.0100$&{1.0100} \\
			{1.00}&$-0.283467$&$0.0000$&{1.0000} \\
		\end{tabular}
	\end{ruledtabular}
\end{table}

{In order to continue, we now need to fix an specific expression for the potential $U(h)$ itself. We then adopt an expression similar to the one which appears in the asymptotic behavior exposed in the previous Eq. (\ref{uh0}), that is
\begin{equation}
U(h) =4h^2, \label{gobsp}
\end{equation}%
which is a power of the so-called old baby Skyrme potential $U_{o}(h)=2h$. Then the superpotential equation (\ref{vvrs}) reads
\begin{equation}
\frac{\lambda ^{2}}{8}\left( {\frac{d\mathcal{W}}{dy}}\right) ^{2}+\left[
\frac{\lambda ^{4}g^{2}}{2}\left( \gamma -h^{2}\right) ^{\beta }\mathcal{W}%
^{2}-4\mu ^{2}h^{2}\right] \left( {\frac{dh}{dy}}\right) ^{2}=0\text{,}
\label{bpsx3}
\end{equation}
in terms of the coordinate $y=r^{2}/2$ and considering $\mathcal{W}_{y}= \mathcal{W}_{h}h_{y}$.}

{Therefore,  the Eq. (\ref{bpsx3}) and the BPS ones (\ref{bpsx1}) and (\ref{bpsx2} constitute a system of differential equations which must be solved numerically according to the boundary conditions (\ref{bc2}), (\ref{bc3}) and (\ref{bbccW}) in terms of the $y$-variable.}

{Below, to compare the numerical results with those obtained via the analytical superpotential, we again fix  $\beta =2$, $g=1$, $\lambda =1$, and  $N=1$. Moreover, we set $\mu^{2}=0.1$.}

{{Figure 4 shows} the numerical solutions to both $h(y)$(top) and $a(y)$ (bottom). We see that these profiles behave in the same general way as in the previous case, including the arising of plateaus (which, as before, can be understood as the origin of the formation of internal structures that characterize the solution to the magnetic sector) in the solutions to the gauge profile function for intermediary values of $y$.}

{{The numerical solutions shown in Figure 5 depict the BPS magnetic field $B(y)$.} Again, the numerical behavior mimics the one obtained previously (i.e., for a purely analytical superpotential), including the presence of internal structures for intermediary $y$. In particular, the value of $B_{0}$ is controlled by $\gamma$ in the very same way as before, such as can be seen from the Eq. (\ref{bpsx1})} {at $y=0$,
\begin{equation}
B_{0}=\mp \lambda ^{2}g^{2}\left( \gamma -1\right) ^{\beta }\mathcal{W}_{0}.
\end{equation}%
In this regard, the magnetic field at the origin always vanishes for $\gamma =1$, independently of the value for $\mathcal{W}_{0}$.}

\vspace{0.75cm}

{\subsection{Additional considerations}}

{We now consider some attributes of our generalized model beyond its BPS framework and the corresponding solutions. {This section aims to demonstrate that the standard Skyrme-Maxwell scenario's basic properties remain unaltered when adding a magnetic permeability. In order to perform such goal,} we follow the prescription previously stated in Ref. \cite{a1} for the Skyrme-Maxwell model.}

\vspace{0.5cm}

\subsubsection{{On the existence of BPS solutions}}

{We first consider the superpotential equation (\ref{Vv}), where $V$, $G$, and $\mathcal{W}$ are functions of $h$ only, i.e. (here, $\mathcal{W}_{h}=d\mathcal{W}/dh$)}
\begin{equation}
V=\frac{\lambda ^{4}g^{2}}{2G}\mathcal{W}^{2}+\frac{\lambda ^{2}}{8}\left(\mathcal{W}_{h}\right) ^{2}\text{,}  \label{rsvv}
\end{equation}%
{which we solve for $\mathcal{W}_{h}$ as}
\begin{equation}
\mathcal{W}_{h}=\sqrt{\frac{8}{\lambda ^{2}}V-4g^{2}\lambda ^{2}\frac{%
\mathcal{W}^{2}}{G}}\text{,}  \label{rsvv1}
\end{equation}%
{from which we calculate $\mathcal{W}_{hh}$, i.e.}
\begin{equation}
\mathcal{W}_{hh}=\frac{4}{\mathcal{W}_{h}}\left[ \frac{1}{\lambda ^{2}}%
V_{h}-g^{2}\lambda ^{2}\frac{\mathcal{W}}{G}\left( \mathcal{W}_{h}-\frac{%
\mathcal{W}}{2G}G_{h}\right) \right] \text{,}
\end{equation}%
{which reveals that $\mathcal{W}_{h}=0$ leads to a nonsingular $\mathcal{W}_{hh}$ only provided that $V_{h}=0$ and $G_{h}=0$. In other words, if $V_{h}\neq 0$ or $G_{h}\neq 0$, $\mathcal{W}_{h}=0$ produces a singularity, and therefore a regular superpotential cannot be defined within the target space, from which we conclude that the corresponding theory does not support BPS solutions. As a consequence, also the \textit{Conjecture 2} as stated in the Ref. \cite{a1} continues to hold even in the enlarged scenario defined in terms of a nontrivial permeability, i.e., BPS solitons exist if and only if the superpotential equation admits a well-defined solution on the whole target space, with $\mathcal{W}_{h}=0$ in the corresponding open interval.}

\begin{figure}[tbp]
\includegraphics[width=8.4cm]{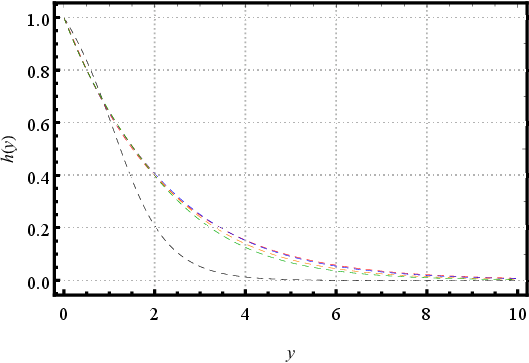}\vspace{0.5cm} %
\includegraphics[width=8.4cm]{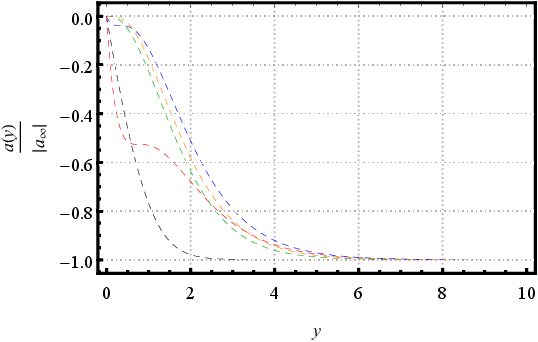}
\caption{Numerical solutions to both $h(y)$ (top) {and }$a(y)$ (bottom,
depicted in units of $\left\vert a_{\infty }\right\vert $) obtained via the
first-order system formed by the Eqs. (\ref{bpsx1}) and (\ref%
{bpsx2}) and (\ref{bpsx3}) for $\gamma =0.50$\ (red line), $%
\gamma =0.75$ (blue line), $\gamma =0.90$ (orange line) and $%
\gamma =1.00$ (green line). The corresponding usual profile (obtained via $%
G=1$) appears as the black line, for the sake of comparison.}
\end{figure}

{As before, one can always imagine the existence of a particular point $h=h_{s}$ within the target space at which $\mathcal{W}_{h}(h=h_{s}) =0$, $V_{h}(h=h_{s})=0$, and  $G_{h}(h=h_{s})=0$ simultaneously as an exception to the aforecited Conjecture 2. In such a case, a {well-defined superpotential $\mathcal{W}(h)$} can be obtained from the superpotential equation. However, given $\mathcal{W}_{h}=0$, the BPS Eq. (\ref{be2}) (which does not depend on the magnetic permeability explicitly, i.e. is the very same one that appears in the standard case) predicts a Skyrme profile function $h(r)$ with a nonmonotonic behavior. The question here is that this same argument can be applied to any arbitrary point $h=h_{s}$. In this sense, for $h_{s}$  sufficiently close to $1$, the corresponding solution reaches values that are greater than the unity and therefore are outside the target space, which is incompatible with the boundary conditions $h(r=0)=1$ and $h(r\rightarrow \infty)\rightarrow 0$.}


\subsubsection{{Bogomol'nyi bound for $g\rightarrow 0$}}

{In Ref. \cite{a1}, the authors argued that potentials of the type $V\sim h^\alpha$ (with $\alpha >0$) allow for the existence of a global (i.e. defined in the whole target space) {solution $\mathcal{W}(h)$ for the standard superpotential equation and, therefore, for the complete BPS scenario (with a Bogomol'nyi bound and BPS solutions). This way, the point to be enlightened is that if a BPS bound exists, it must attain the same value} inherent to the ungauged BPS baby Skyrme model in the limit of a vanishing electromagnetic coupling constant $g$. }

\begin{figure}[tbp]
\includegraphics[width=8.4cm]{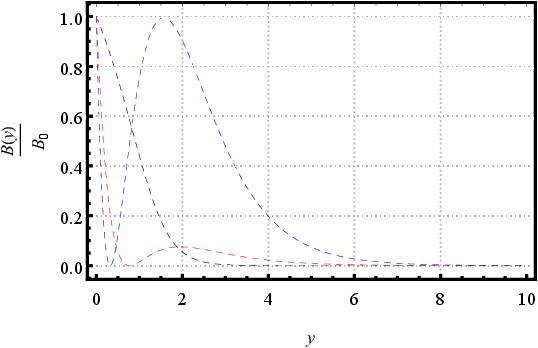}\vspace{0.5cm} %
\includegraphics[width=8.4cm]{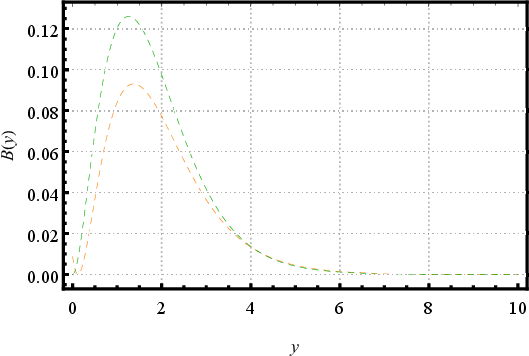}
\caption{Numerical solutions to the BPS magnetic field $B(y)$ obtained via
the first-order system formed by the Eqs. (\ref{bpsx1}) and (\ref{bpsx2}) and (\ref{bpsx3}). Conventions as in the Fig. 1. Top: the profiles (depicted in units of $B_{0}$) for $\gamma =0.50$ and $\gamma =0.75$. Bottom: the results for $\gamma =0.90$ and $\gamma =1.00$. The corresponding standard solution again appears as the black line. Here, some of the noncanonical profiles were normalized for the sake of visualization.}
\end{figure}

{We now verify whether such a convergence {still holds} even in our generalized case. To attain this goal, we implement the prescription established by the Ref. \cite{a1} and expand the second root, which appears on the right-hand side of the Eq. (\ref{rsvv1}) in a power series in $g^{2}$, {via which, by considering only the first relevant terms, we obtain}}
\begin{equation}
\mathcal{W}_{h}\approx \frac{2}{\lambda }\sqrt{2V}\left( 1-\frac{%
g^{2}\lambda ^{4}}{4}\frac{\mathcal{W}^{2}}{VG}\right) .\label{wwhh}
\end{equation}%

{In view of the expansion, we also consider the superpotential at second-order in $g$, i.e.}
\begin{equation}
\mathcal{W}\approx \mathcal{W}^{(0)}+g^{2}\mathcal{W}^{(2)}\text{,}  \label{asw}
\end{equation}%
{from which the Eq. (\ref{wwhh}) provides a set of differential equations for every order in $g$ as}
\begin{equation}
\frac{\partial \mathcal{W}^{(0)}}{\partial h}=\frac{2\sqrt{2}}{%
\lambda }\sqrt{V}\text{,}  \label{fe1}
\end{equation}%
\begin{equation}
\frac{\partial \mathcal{W}^{(2)}}{\partial h}=-\frac{\sqrt{2}%
\lambda ^{3}}{2}\frac{\left( \mathcal{W}^{\left( 0\right) }\right) ^{2}}{G%
\sqrt{V}}\text{.}  \label{fe2}
\end{equation}

{In the sequence, by regarding the family of potentials $V(h)=h^{2\alpha}$ ($\alpha >0$), we promptly integrate the Eq. (\ref{fe1}) and obtain}
\begin{equation}
\mathcal{W}^{(0)}=\frac{2\sqrt{2}}{\lambda \left( \alpha
+1\right) }h^{\alpha +1}\text{,}  \label{fe3}
\end{equation}%
{via which the Eq. (\ref{fe2}) assumes the form}
\begin{equation}
\frac{\partial \mathcal{W}^{(2)}}{\partial h}=-\frac{4\sqrt{2}%
\lambda }{\left( \alpha +1\right) ^{2}}\frac{h^{\alpha +2}}{G}. \label{w22}
\end{equation}

{The equation above clarifies that the solution for $\mathcal{W}^{(2)}$ depends upon the magnetic permeability $G$, i.e. according to Eq. (\ref{asw}), the superpotential $\mathcal{W}(h)$  echoes the presence of $G(h)$ starting from the second-order in the electromagnetic coupling constant. In what follows, we also consider  $G(h)=(\gamma -h^{2})^{-2}$, from which we write Eq. (\ref{w22}) as}
\begin{equation}
\frac{\partial \mathcal{W}^{(2)}}{\partial h}=-\frac{4\sqrt{2}%
\lambda }{\left( \alpha +1\right) ^{2}}h^{\alpha +2}\left( \gamma
-h^{2}\right) ^{2}\text{,}
\end{equation}%
{whose solution reads}%
\begin{equation}
\mathcal{W}^{(2)}=-\frac{4\sqrt{2}\lambda }{\left( \alpha
+1\right) ^{2}}h^{\alpha +3}\left( \frac{\gamma ^{2}}{\alpha +3}-\frac{%
2\gamma h^{2}}{\alpha +5}+\frac{h^{4}}{\alpha +7}\right) \text{,}
\end{equation}%
{which, together with the Eq. (\ref{fe3}), leads to the solution for the superpotential in the limit of sufficiently small $g$, i.e.}%
\begin{eqnarray}
&&\left. \mathcal{W}(h)=\frac{2\sqrt{2}h^{\alpha +1}}{\lambda \left( \alpha
+1\right) }\left[ 1-\frac{2\lambda ^{2}g^{2}h^{2}}{\alpha +1}\left( \frac{%
\gamma ^{2}}{\alpha +3}\right. \right. \right.   \notag \\
&& \left. \left. -\frac{2\gamma h^{2}}{\alpha +5}+\frac{h^{4}}{\alpha +7}%
\right) \right].  \label{aw2}
\end{eqnarray}
{{The evaluation of the above expression at $h=1$ provides the BPS bound, which allows us to} conclude that our enlarged scenario correctly reproduces the bound inherent to the ungauged baby Skyrme case (the interested reader may compare the above expression with the Eq. (112) of the Ref. \cite{a1}). The novelty here appears in the leading correction for small $g$ (which is of order $g^{2}$ and negative, as in the Skyrme-Maxwell case with $G=1$), i.e. a magnetic permeability affects not only the general behavior of the corresponding term (through the power of $h$) but also its value calculated at $h=1$.}


\subsubsection{{Magnetic flux}}

{It is also interesting to clarify whether the presence of a nontrivial magnetic permeability affects the value of the magnetic flux calculated for small and large electromagnetic coupling $g$. In order to offer a response to this question, we first observe that the magnetic flux can be expressed as}
\begin{equation}
\Phi _{B}=2\pi \int B(r)rdr=-2\pi Na_{\infty}\text{,}  \label{mf0}
\end{equation}%
{where we have used both the Eq. (\ref{mf}) for the  magnetic field and the conditions (\ref{bc2}). Here, we have defined $a_{\infty}=a(r\rightarrow
\infty )$.}

{In the sequence, we divide the Eq. (\ref{be1}) by the Eq. (\ref{be2}%
), from which we obtain}%
\begin{equation}
{\frac{da}{\left( 1+a\right) }=g^{2}\lambda^2 F_h\, dh}\text{,}\label{fe0}
\end{equation}%
{where}%
\begin{equation}
F_h=\frac{4\mathcal{W}}{\mathcal{W}_{h}G}\text{.}  \label{fh}
\end{equation}

{The Eq. (\ref{fe0}) has the solution (here, }$C${\ stands for
an integration constant)}%
\begin{equation}
\ln \left[ C\left( 1+a\right) \right] =g^{2}\lambda ^{2}F(h)\text{,}
\label{fe0s}
\end{equation}%
{where}%
\begin{equation}
F(h)=4\underset{0}{\overset{h}{\int }}\frac{\mathcal{W}\left( h^{\prime
}\right) }{\mathcal{W}_{h}\left( h^{\prime }\right) G\left( h^{\prime
}\right) }dh^{\prime }\text{.}  \label{df}
\end{equation}

{In order to calculate the value of }$C${, we evaluate the
solution Eq. (\ref{fe0s}) at }$r=0${, which leads to}%
\begin{equation}
C=e^{g^{2}\lambda ^{2}F(1)}\text{,}  \label{sC}
\end{equation}%
{where we have used the boundary conditions }$h(r=0)=1${\ and }%
$a(r=0)=0${.}

\begin{figure}[tbp]
\includegraphics[width=8.4cm]{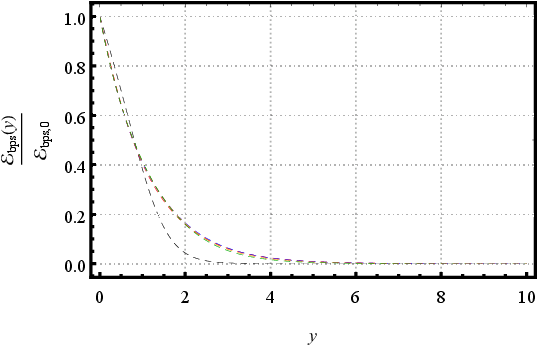}\vspace{0.5cm}
\caption{Numerical solutions to the energy density of the BPS
configurations, i.e. $\varepsilon _{bps}(y)$ (depicted in units of $%
\varepsilon _{bps,0}$). Conventions as in the Fig. 4. The profiles
were normalized for the sake of visualization. }
\end{figure}

{Moreover, at the vacuum $h=0$, the Eq. (\ref{bbccW}) predicts $\mathcal{W}(h=0)=0$, from which it is reasonable to suppose that the potential behaves $V\sim h^{2\alpha}$ (again with $\alpha >0$). In such a scenario, the superpotential equation Eq. (\ref{rsvv}) suggests the adoption of $\mathcal{W}_{h}\sim h^{\alpha }$, $\mathcal{W}\sim h^{\alpha +1}$, and $G^{-1}\sim \Delta +\mathcal{O}\left( h^{2}\right)$ (with $\Delta$ constant;
for $G(h)=\left( \gamma -h^{2}\right)^{-2}$, one gets $\Delta=\gamma ^{2}$), from which the Eq. (\ref{fh}) leads to}
\begin{equation}
F_{h}\sim 4\Delta h\text{,}  \label{fh1}
\end{equation}%
{which indicates that not only $F_{h}$, but also $F(h)$ itself vanishes at $h=0$.}

{Now, whether we evaluate the Eq. (\ref{fe0s}) in the
asymptotic region $r\rightarrow \infty $, we get that (here, we
have used both }$F(h=0)=0${\ and the Eq. (\ref{sC}) for the value of }%
$C$\textbf{)}%
\begin{equation}
a_{\infty }=-1+e^{-g^{2}\lambda ^{2}F(1)}\text{,}
\end{equation}%
{via which the magnetic flux }$\Phi _{B}${\ Eq. (\ref{mf0})
assumes the form}%
\begin{equation}
\Phi _{B}=2\pi N\left[ 1-e^{-g^{2}\lambda ^{2}F(1)}\right] \text{,}
\end{equation}%
{which leads to the following expressions}%
\begin{equation}
g\text{ small: }\Phi _{B}\sim 2\pi Ng^{2}\lambda ^{2}F(1)\text{,}
\label{mfsg}
\end{equation}%
\begin{equation}
g\text{ large: }\Phi _{B}\sim 2\pi N\text{,}
\end{equation}
{{which are the same} ones that appear in the restricted baby Skyrme-Maxwell theory (with $G=1$), see the eqs. (121) and (122) of the Ref. \cite{a1}, respectively. However, despite the same \textit{symbolic} form, the value of the magnetic flux for small $g$ in Eq. (\ref{mfsg}) is now influenced by the nontrivial magnetic permeability via the definition of $F(h)$ given in Eq. (\ref{df}).}


\subsubsection{{Some exact BPS solutions}}

{We end this manuscript by exploring whether our generalized model admits exact solutions. In this sense, we use the prescription proposed in the Ref. \cite{a1} in the context of the canonical Skyrme-Maxwell scenario, the
starting-point being the BPS Eqs. (\ref{be1}), (\ref{be2}), now rewritten
in terms of the variable $y=r^{2}/2$ as}
\begin{equation}
Na_{y}=-\lambda ^{2}g^{2}\frac{\mathcal{W}}{G}\text{,}  \label{ay0}
\end{equation}%
\begin{equation}
4Nh_{y}\left( 1+a\right) =-\mathcal{W}_{h}\text{,}
\end{equation}%
{for $N>0$}

{Whether we introduce the new field $p=h^{2}$, the above
first-order expressions can be written as}
\begin{equation}
Na_{y}=-\lambda ^{2}g^{2}\frac{\mathcal{W}}{G}\text{,}  \label{ay}
\end{equation}%
\begin{equation}
Np_{y}\left( 1+a\right) =-\mathcal{W}_{p}p\text{,}  \label{py0}
\end{equation}%
{where both $\mathcal{W}$ and $G$ are now functions of $p$.}

{In addition, from the Eq. (\ref{ay}), one gets that}%
\begin{equation}
-\frac{N}{\lambda ^{2}g^{2}}a_{yy}=\frac{p_{y}}{G}\left( \mathcal{W}_{p}-%
\frac{\mathcal{W}}{G}G_{p}\right) \text{,}
\end{equation}%
{which can be combined with the eqs. (\ref{ay}) and (\ref{py0}) in
order to give}
\begin{equation}
-Na_{yy}=\frac{a_{y}}{\left( 1+a\right) }\frac{\mathcal{W}_{p}p}{\mathcal{W}}%
\left( \mathcal{W}_{p}-\frac{\mathcal{W}}{G}G_{p}\right) \text{.}
\label{ayy}
\end{equation}

{In order to continue with our construction, it is now necessary to specify both $G(p)$ and $\mathcal{W}(p)$. Then, for such an aim, we set the superpotential as (with $\sigma\geq 1$)
\begin{equation}
\mathcal{W}( p)=\mathcal{W}_{0}p^{\sigma },\label{gpm0}
\end{equation}
where $\mathcal{W}_{0}=\mathcal{W}(p(0))=\mathcal{W}(1)$. In addition, for the magnetic permeability, we choose
\begin{eqnarray}
G( p) &=&G_{0}p^{1-\beta }\;\text{ (for $\sigma=1$)},\label{gpm1}\\[0.2cm]
G( p) &=&G_{0}p^{\sigma}\exp\left(-\frac{\beta(p^{1-\sigma}-1)}{ \sigma(1-\sigma)}\right)\;\text{(for $\sigma> 1$)},\quad \label{gpm2}
\end{eqnarray}
with $G_{0}=G(p(0))=G(1)$. Here, we also have defined the parameter $\beta$ as
\begin{equation}
\beta =\frac{\delta }{\mathcal{W}_{0}}.
\end{equation}

By substituting both the equations above in Eq. (\ref{ayy}), we attain a differential equation similar to that found in Ref. \cite{a1}, i.e.
\begin{equation}
-Na_{yy}=\delta\,\frac{ a_{y}}{1+a},\label{ayy1}
\end{equation}
whose solution for the gauge profile function $a(y)$ reads
\begin{equation}
a(y) =-1+\frac{1}{C}\text{Li}^{-1}\left( \text{Li}(C) -\frac{C\delta }{N}y\right), \label{aayext}
\end{equation}
expressed in terms of the logarithmic integral function \textrm{Li}, where the parameter $C$ now is read
\begin{equation}
\quad C=\exp \left( \frac{\lambda ^{2}g^{2}\mathcal{W}_{0}}{G_{0}\delta}\right).
\end{equation}}

{Immediately, we also solve the Eq. (\ref{ay0}), which provides the following expressions for $h(y)=\sqrt{p(y)}$, i.e.
\begin{equation}
h( y)  =\left( \frac{\ln[ C( 1+a(y))]}{\ln(C)}\right)^{\frac{1}{2\beta}},
\end{equation}
which holds for $\sigma =1$, while
\begin{equation}
h( y)=\left[ 1+\frac{\sigma( 1-\sigma)}{%
\beta }\ln \left( \frac{\ln \left[ C\left( 1+a\left( y\right) \right) \right]
}{\ln \left( C\right) }\right) \right] ^{\frac{1}{2(1-\sigma)}},
\end{equation}%
holds when $\sigma >1$. Here, $a(y)$ is given by the Eq. (\ref{aayext}).}

{The solutions for $\sigma\geq 1$ are extended profiles along all the radial axis. Thus, the behavior in the limit $y\rightarrow \infty$ for the gauge field reads
\begin{equation}
a( y)\thickapprox a_\infty+\frac{1}{C} \exp\left(-\frac{C\delta}{N}y \right),
\end{equation}
where $a_\infty=-1+C^{-1}$. We therefore observe that the respective tail follows a Gaussian-law decay. On the other hand, the behavior of the Skyrmion profile function depends on the values of $\sigma$. This way, for $\sigma=1$, such a function also presents a Gaussian-law decay, i.e.
\begin{equation}
h( y)\thickapprox \left( \frac{1}{\ln \left( C\right) }\right) ^{\frac{1}{%
2\beta }}\exp \left( -\frac{C\delta }{2\beta N}y\right),
\end{equation}
while for $\sigma>1$, the respective tail follows a power-law decay,
\begin{equation}
h( y) \approx \left( \frac{\beta N}{\delta C\sigma(\sigma -1)}\right)^{\frac{1} {2( \sigma -1)}}y^{-\frac{1}{2( \sigma -1) }}.
\end{equation}}

\begin{table}[t]
\caption{Approximate numerical values used for the normalization of the noncanonical profiles obtained from the Eqs. (\ref{bpsx1}), (\ref{bpsx2}),  and (\ref{bpsx3}), for $\beta =2$, $g=1$, $\lambda =1$, $N=1$ and $\mu ^{2}=0.1$. The standard values (with $\beta =0$) are $a_{\infty}= -0.679311$, $B_{0}=0.7057$ and $\varepsilon _{bps,0}=0.8$, for the sake of comparison.}
\begin{ruledtabular}
		\begin{tabular}{cccc}
	$\gamma$& $\quad a_{\infty}$ &$B_{0}$&$\varepsilon _{bps,0}$\\
			\colrule
			 & & & \\ [-0.25cm]	
			{0.50}&$-0.082308$&$0.2189$&{0.8000} \\
			{0.75}&$-0.136245$&$0.0551$&{0.8000} \\
			{0.90}&$-0.217839$&$0.0087$&{0.8000} \\
			{1.00}&$-0.286781$&$0.0000$&{0.8000} \\
		\end{tabular}
	\end{ruledtabular}
\end{table}

{It is important to note that, {given our choices (\ref{gpm0}), (\ref{gpm1}), and (\ref{gpm2}) for $\mathcal{W}(p)$ and $G(p)$,} the factor $\mathcal{W}^{-1} \mathcal{W}_{p} p \left(\mathcal{W}_{p}-G^{-1} \mathcal{W}G_{p}\right)$ which appears in the Eq. (\ref{ayy}) can be reduced to a constant (in this case, $\delta$). As a direct consequence, we arrive at the Eq. (\ref{ayy1}) (containing $a(y)$ itself and its derivatives only), which also plays a central role in the construction of our exact solutions.}


\section{Summary and perspectives\label{sec4}}

{We have investigated BPS solitons inherent to a gauged baby Skyrme scenario {immersed in a magnetic medium.} We have minimized the corresponding {total energy by implementing the BPS technique, from which we have verified that} the enlarged model also possesses a well-defined BPS structure. {As expected, it allows us to attain the self-dual} equations and a lower bound for the total energy. In such a context, we have clarified how the permeability enters {the differential relation between the superpotential and the corresponding BPS potential, and also the self-dual equation which defines the magnetic field. Consequently, the permeability may engender a magnetic field with an internal structure, i.e. one that behaves in the standard way at the vacuum but sometimes not at the origin. Besides, along the radial axis, its profile can have a format different from the one found in the canonical gauged BPS Skyrme model.}}

{To solve the BPS system of differential equations, we have focused our attention on those configurations possessing a rotational symmetry described by the profile functions $h(r)$ and $a(r)$. Next, after choosing an analytical expression for the permeability function (which includes the parameter $\gamma\in \mathds{R} $), we have explored two different scenarios based on the nature of the superpotential: In the first case, we set the superpotential as an explicit function of $h(r)$, such that, together with the permeability, the BPS potential be defined analuytically. In contrast, in the second situation, we fix the particular form of the BPS potential and determine the superpotential through the numerical solution of a system of differential equations formed by the two BPS equations and the superpotential equation itself.}

{{For the sake of comparison, in both scenarios, we have worked with potentials whose behaviors when approaching the vacuum are similar (here, both potentials behave as $h^{2}$). We have then numerically solved the two scenarios and depicted the corresponding profiles in the figs. 1, 2 and 3 (for the analytical superpotential), and in the figs. 4, 5 and 6 (for the second case).} The resulting solutions have revealed how the profiles change with variations on $\gamma$, giving rise to configurations with internal structures.} {In particular, for the case with an exact superpotential, we have {analyzed the main features inherent to the shape of the magnetic field emerging within a range of values of $\gamma$.} This study has shown, for instance, the relation between the amplitudes of the peaks {(local maxima) inherent to the magnetic profile, showing a format which differs dramatically from the one obtained in the gauged BPS Skyrme model. Similarly, we could analyze the form of the magnetic field for other} values of $\gamma$.}

{Beyond the BPS framework and its solutions, we have investigated some basic properties of our enlarged Skyrme-Maxwell scenario. For instance, we have verified that it mimics {some properties which the canonical model itself satisfies.} In particular, we have clarified that both the \textit{Conjecture 1} and \textit{Conjecture 2} and the \textit{Corollary 1} stated in the Ref. \cite{a1} continue to hold. {As a second point, we have also discovered that our generalized theory correctly reproduces the bound inherent to the ungauged baby Skyrme model in the limit of a sufficiently small electromagnetic coupling constant $g$. In addition, the effects of our generalization also appear at $g^{2}$-order.}} {Subsequently, in our third item, we have studied how permeability affects the value of the magnetic flux $\Phi _{B}$ for both small and large values of $g$ by obtaining the very same [symbolic] analytical expressions that appear in the standard case. Thus, for small $g$, the permeability influences the magnetic flux. In contrast, for large $g$, the flux $\Phi _{B}$ remains unaltered. Finally, we have selected a class of functions defining the superpotential and magnetic permeability which generate a family of noncompact solutions. Whereas the gauge field and the magnetic one attain their vacuum values by following a Gaussian-law decay, the tail of the Skyrmion's profile follows a Gaussian-law decay or a power-law one.}

{The results introduced in this manuscript aim to contribute to the understanding of the electromagnetic properties of gauged Skyrmions by studying its gauged BPS baby Skyrme versions. We now intend to apply the same approach to other gauged versions of the restricted baby Skyrme enlarged model, for instance, in the presence of the Chern-Simons' or Born-Infeld's gauge fields. The results concerning these perspectives will be reported in future contributions.}

\begin{acknowledgments}
The authors thank Prof. Dion\'{\i}sio Bazeia and Prof. Lukasz Stepien for motivating discussions. This work was financed in part by the Coordena\c{c}\~{a}o de Aperfei\c{c}oamento de Pessoal de N\'{\i}vel Superior - Brasil (CAPES) - Finance Code 001, the Conselho Nacional de Pesquisa e Desenvolvimento Cient\'{\i}fico e Tecnol\'{o}gico - CNPq and the Funda\c{c}\~{a}o de Amparo \`{a} Pesquisa e ao Desenvolvimento Cient\'{\i}fico e Tecnol\'{o}gico do Maranh\~{a}o - FAPEMA (Brazilian agencies). {In particular, J. A.  thanks the full support from CAPES (via a PhD scholarship). R. C. acknowledges the support from the grants CNPq/306724/2019-7, CNPq/312155/2023-9,  FAPEMA/Universal-01131/17, FAPEMA/Universal-00812/19 and FAPEMA/APP-12299/22. E. H. thanks the support from the grant CNPq/309604/2020-6. A. C. S. thanks the grants CAPES/88882.315461/2019-01 and CNPq/150402/2023-6.}
\end{acknowledgments}

\end{document}